\documentclass[]{revtex4}
\usepackage{ifpdf}
\usepackage{color}
\usepackage{graphicx}  
\usepackage{epstopdf}

\usepackage{float}
\usepackage{amsmath}
\usepackage{acronym}
\usepackage{multirow}
\usepackage{xspace}
\usepackage{units}
\usepackage{hyperref}
\usepackage{amsfonts}
\usepackage{bm}
\usepackage{braket}

\def\thercsid{\relax}
\def\rcsid#1{\def\next##1#1{\def\thercsid{##1}}\next}
\rcsid$Id: s6pe.tex,v 1.216 2013/03/18 21:12:42 vivien Exp $
\renewcommand{\today}{\number\day\space\ifcase\month\or
  January\or February\or March\or April\or May\or June\or
  July\or August\or September\or October\or November\or December\fi
  \space\number\year}

\newcommand{\LALInference}{\texttt{LALInference}}
\newcommand{\BayesWave}{\texttt{BayesWave}}
\newcommand{\BayesLine}{\texttt{BayesLine}}

\newcommand{\data}{\ensuremath{\bm{d}}}
\newcommand{\model}{\ensuremath{\mathcal{M}}}

\renewcommand{\vec}[1]{\ensuremath{\bm{#1}}}
\newcommand{\pvec}{\ensuremath{\bm{\theta}}}

\newacro{BNS}{binary neutron star}
\newacro{NSBH}{neutron star -- black hole binary}
\newacro{BBH}{binary black hole}
\newacro{SNR}{signal-to-noise ratio}
\newacro{PDF}{probability density function}
\newacro{PSD}{power spectral density}
\newacro{GW}{gravitational wave}
\newacro{CBC}{compact binary coalescence}

\begin{document}
\title{\BayesLine: Bayesian Inference for Spectral Estimation of Gravitational Wave Detector Noise }

\author{Tyson B. Littenberg}
\address{Center for Interdisciplinary Exploration and Research in
Astrophysics (CIERA) \& Department of Physics and Astronomy, Northwestern University,
2145 Sheridan Road, Evanston, IL 60208}
\author{Neil J. Cornish}
\address{Department of Physics, Montana State University, Bozeman,
MT 59717, USA}

\date{\today}

\begin{abstract}

Gravitational wave data from ground-based detectors is dominated by instrument noise. Signals will be comparatively weak, and our understanding of the noise will influence detection confidence and signal characterization. Mis-modeled noise can produce large systematic biases in both model selection and parameter estimation. Here we introduce a multi-component, variable dimension, parameterized model to describe the Gaussian-noise power spectrum for data from ground-based gravitational wave interferometers. Called BayesLine, the algorithm models the noise power spectral density using cubic splines for smoothly varying broad-band noise and Lorentzians for narrow-band line features in the spectrum. We describe the algorithm and demonstrate its performance on data from the fifth and sixth LIGO science runs. Once fully integrated into LIGO/Virgo data analysis software, BayesLine will produce accurate spectral estimation and provide a means for marginalizing inferences drawn from the data over all plausible noise spectra.
\end{abstract}

\maketitle

\section{Introduction} \label{intro}

Gravitational wave data from ground based observatories such as LIGO~\cite{LIGO} and Virgo~\cite{Virgo} are dominated by instrument noise.  Even with the improved hardware capabilities of the Advanced LIGO/Virgo interferometers~\cite{aLIGO,aVirgo} sensitive analysis methods must be in place to take full advantage of the data.  While a wide variety of search pipelines exist for a similarly wide variety of GW source types (e.g.~\cite{Abbott:2009up,Abadie:2010yb,StochasticS5}), all fundamentally stem from the same principle -- assessing whether the detector output is statistically different from our model for the data.  

For analyses which have the advantage of strong theoretical predictions for the gravitational wave signal, such as searches for the inspiral and merger of stellar mass compact binaries (neutron stars and/or black holes) template based techniques are optimal~\cite{Finn:1992wt,Cutler:1992tc}.  In the template based analyses used for signal characterization~\cite{S6PE,LALInference}, a trial waveform, or ``template,'' is subtracted from the data and the residual is compared to a theoretical model for the instrument noise.  The statistic used to assess whether the residual and the noise model are consistent is the likelihood, or the probability density of the detectors producing data $d$ for a hypothetical gravitational wave signal $h$ with parameters $\bf{\theta}$.  Assuming we understand the properties of the noise distribution, the likelihood will be maximized when the template waveform matches the true signal modulo statistical fluctuations due to the specific realization of the noise.

Any unsound approximations in the likelihood will contribute to systematic errors which could result in erroneous inferences as the template waveform parameters flex away from the ``true'' values in an attempt to achieve a the expected statistics for the residual.  Receiving most of the spotlight in the effort to control systematic errors are the template waveforms themselves~\cite{Canitrot:2001hc,Ajith:2009fz,Buonanno:2007pf,Littenberg:2012uj,Favata:2013rwa}.  However, understanding the noise is just as important if we want to ensure accurate astrophysical parameter estimation and discoveries, or limits pertaining to fundamental physics such as alternative theories of gravity~\cite{Cornish:2011ys,Li:2011cg,Sampson:2013jpa} and the equation of state of neutron stars~\cite{Flanagan:2007ix,Read:2009yp,Lackey:2013axa,Wade:2014vqa}.

The LIGO/Virgo noise is dominated by three main components:  Seismic noise steeply limits sensitivity below 10 Hz; thermal noise from the mirror suspensions and coatings dominate between 10 and 200 Hz; and quantum (photon shot) noise is the limiting factor at high frequencies.  Each component is a broad-band effect with relatively simple frequency dependence.  Standing out above the broad-band noise are high power, narrow band, spectral lines which originate from a variety of sources including the mirror suspensions, the AC electrical supply, or sinusoidal motion imparted on the mirrors for calibration.

Approaches for improving the noise model include modifying the functional form of the likelihood~\cite{Allen:2002jw,Rover:2008yp,Rover:2011qd}; demanding additional consistency checks between analyses on single detectors and the network~\cite{Veitch:2008ur,Veitch:2009hd}; or introducing degrees of freedom which help fit to the instrument noise~\cite{Littenberg:2010gf,Littenberg:2013gja}. Here we introduce a new way of parameterizing LIGO/Virgo Gaussian noise and use modern data analysis methods for fitting to, and eventually marginalizing over, the noise model.  A companion publication will address the non-stationary and non-Gaussian noise by introducing the \BayesWave\ algorithm~\cite{BayesWave}; a trans-dimensional Bayesian approach similar in spirit to \BayesLine\ but employing a different basis -- wavelets -- to model transient, non-Gaussian noise.    

In Section~\ref{sec:motive} we motivate the need for real-time spectral estimation, in Section~\ref{sec:PSD} we consider features commonly found in LIGO and Virgo noise spectra which place different demands on our model.  Section~\ref{sec:BayesLine} introduces the \BayesLine\ algorithm and we demonstrate its performance on LIGO data in \ref{sec:S6}.  Our conclusions are summarized in Section~\ref{sec:Conclusions} while Appendix \ref{sec:60Hz}  briefly explores the prospects of coherently fitting out spectral lines associated with the electrical power supply.

\section{Motivation} \label{sec:motive}

The noise from gravitational wave detectors $\vec{n}$ is approximately Gaussian with zero mean (i.e. $\langle n \rangle=0$) and stationary (meaning the noise statistics are not time dependent).  Here, and throughout, bold-faced quantities indicate the vector of all samples, while un-bolded quantities refer to a single sample.  Because the autocorrelation matrix of stationary noise depends only on the lag, the noise covariance matrix of $\tilde{\bf{n}}$, where the tilde represents a Fourier transform, is diagonal (i.e., there are no correlations between noise at different frequencies).  Under these conditions we can completely characterize the noise by its one-sided power spectral density (PSD) $S_n(f)\equiv \frac{2}{T}\langle |\tilde{n}(f)|^2\rangle$ where $T$ is the duration of data being analyzed, typically 10 s to 1000 s  for LIGO/Virgo compact binary signals.  In the paradigm of stationary and Gaussian noise we arrive at the standard likelihood used in gravitational wave data analysis~\cite{Finn:1998vp}

\begin{equation}
\log p\left(\vec{d}\middle|\vec{\theta}\right) = -\sum_{i}^{\rm Net}\left[\int_0^{\infty} \frac{|\tilde{\vec{r}}^i(f;\vec{\theta}) |^2}{S_n^i(f)} df \right] + C
\label{eq:likelihood}
\end{equation}
where $\tilde{\vec{r}}(f;\vec{\theta}) = \tilde{\vec{d}}(f)-\tilde{\vec{h}}(f;\vec{\theta})$ is the residual noise after the gravitational wave template has been regressed from the data, and the summation is over detectors in the network (e.g., the two LIGO detectors and Virgo).

Characterizing the detector noise is a challenge because it is not stationary for times much longer than a few tens of second, depriving us of a sufficiently accurate ``reference" noise spectrum that can be used for all analyses.  Instead, the noise floor must be estimated from the data directly.  Different analyses have different prescriptions for PSD estimation.  Past searches for binary in-spiral signals used a running average of the instrument power spectrum over long durations of time~\cite{Abadie:2010yb} while the parameter estimation follow-up analysis relied on Welch averaging the power spectrum of many segments of data near in time to a candidate signal coming from the search pipeline~\cite{S6PE, LALInference}.  Searches for un-modelled signals have their own methods for PSD estimation, including pre-conditioning the data by filtering narrow-band spectral features and piece-wise whitening the Fourier domain data~\cite{Finn:1999cx}, and whitening across frequency layers in a discrete wavelet transform representation of the data~\cite{Klimenko:2008fu}. 

Consider the parameter estimation follow-up pipeline approach to PSD estimation.  Relying on off-source data to estimate the noise level presents several hazards including times when the detector noise experiences large excursions due to impulsive noise events, ``glitches,'' or long duration fluctuations of data quality due to prolonged changes in the environment such as seismic activity, wind, etc.  In past analyses, times used to produce PSDs were manually inspected to ensure quality data was employed in the Welch averaging and the windows of time were adjusted accordingly.  Finding adequate data for noise estimation could be challenging, and the posterior distribution functions recovered occasionally showed statistically significant dependence to the data used for PSD estimation~\cite{S6PE,Littenberg:2013gja}.  

The challenge of producing good initial estimates of the noise power spectrum will be compounded in the advanced detector era.  As the sensitivity of the detectors improves, particularly at low frequency, the amount of time that a gravitational wave chirp signal will be in band increases.  For example,  a $1.4 - 1.4$ ${\rm M}_\odot$ binary neutron star merger takes $\sim25$ seconds to evolve from 40 Hz (the approximate low frequency cut-off during the initial science runs) to the point where the stars merge.  Advanced LIGO/Virgo are expected to ultimately reach down to frequencies of $\sim10$ Hz~\cite{Waldman:2011}, at which point the binary is in band for $\sim10^3$ seconds.  Reliance on averaging for PSD estimation will demand $\sim10^4$ seconds -- hours -- of stationary noise similar to the data containing the candidate event.  The demands on detector stability for the durations needed to achieve good noise estimates are impractical.

It was demonstrated in Ref.~\cite{Littenberg:2013gja} that sensitivity to the initial PSD estimation could be mitigated, but not eradicated, by marginalizing over the overall power in the instrument background using parameters which produced a step-wise re-scaling of the noise spectrum.  The PSD rescaling is a step in the right direction, and is computationally inexpensive, but requires an initial estimate of the PSD and is limited in its flexibility; it was unable to modify the spectral slope or the location and relative amplitude of the spectral lines found in LIGO/Virgo data.

Given this reality, an alternative approach to determining the PSD -- one which only relies on the data containing the trigger -- is desirable.  Ideally we would have a parameterized model for the instrument noise, just as we do for the gravitational wave signal, and the two will be deduced simultaneously during the parameter estimation analysis.  To that end, we present a novel approach to PSD estimation where we introduce a variable number of phenomenological parameters which are sufficiently flexible to fully characterize the noise power spectrum but are held in check using Bayesian model selection to prevent over-fitting the data.  The PSD model is broken up into two components:  Cubic spline interpolation is used to fit to the broad-band spectrum dominated by seismic, thermal, and shot noise while a linear combination of Lorentzians is used to whiten the narrow-band resonant features in the data due in part to the AC power supply, calibration lines, and vibrational modes of the suspension system supporting the test masses.

\section{The LIGO/Virgo power spectrum}\label{sec:PSD}

\begin{figure}[htbp]
   \centering
   \includegraphics[width=\linewidth]{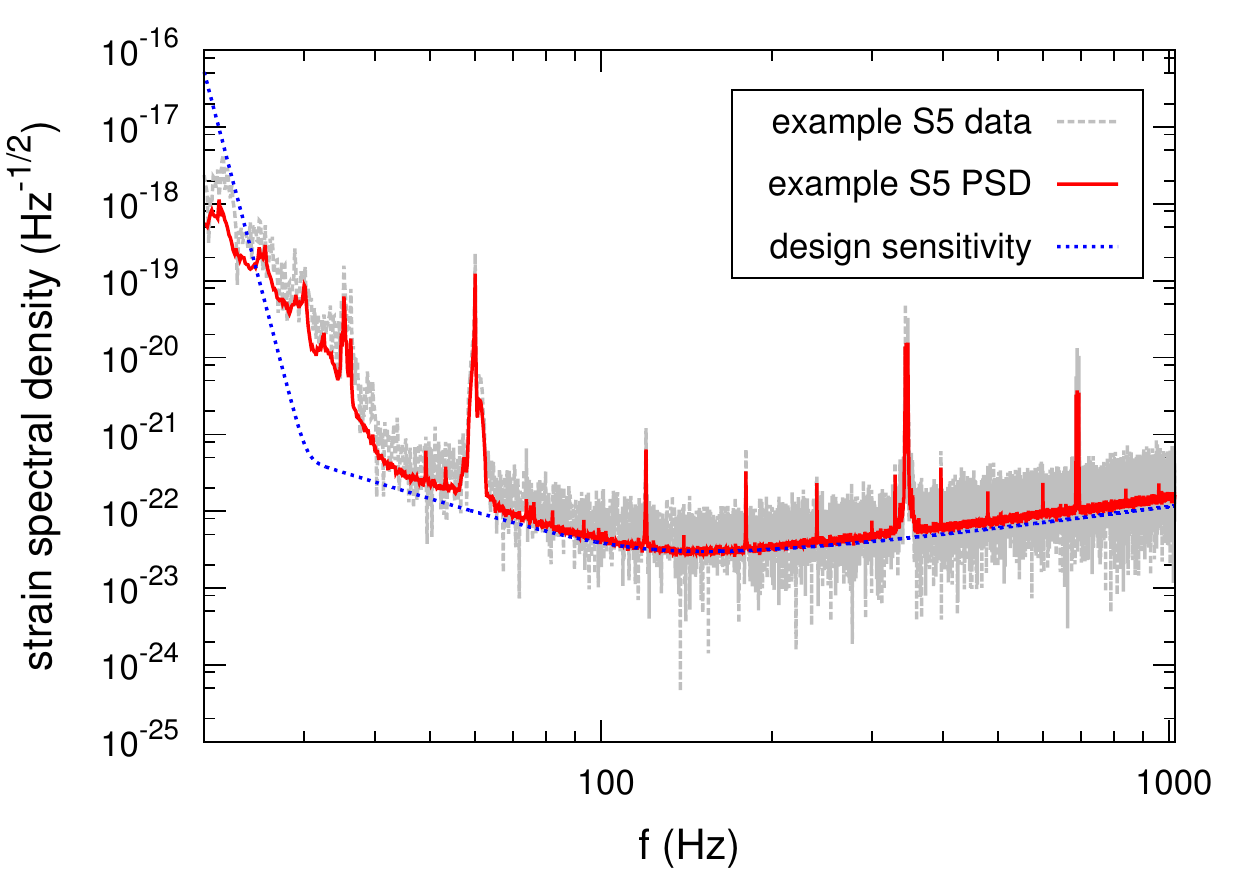} 
   \caption{\small{Example strain spectral density (square root of the power spectrum; gray, dashed) and $\sqrt{2S_n(f)/T}$ (red, solid) during LIGO's $5^{\rm th}$ science run, compared to the design sensitivity (blue, dotted).  Notice that the design sensitive consists strictly of broad-band features due to seismic, thermal, and quantum noise, while the actual data show high-amplitude narrow-band spectral lines.  The design sensitivity is the type of noise curve used for studies in simulated data, while the red (solid) line is needed for analyzing the actual data.}}
   \label{fig:budget}
\end{figure}

Before launching into the details of how we will model the PSD, and how this improvement impacts inferences drawn from the data, we will first motivate the need for more sophisticated treatment of the instrument noise by demonstrating the breakdown of the standard assumptions about the data being stationary on timescales needed to perform in-spiral analyses.  To do so we will use data from the LIGO Livingston Observatory to illustrate the challenges facing future observations.  To start, we will begin with an example power spectrum of typical LIGO data shown in the gray (dashed) line of Figure~\ref{fig:budget}.  Plotted over the spectrum is the design sensitivity (blue, dashed line) which is the type of PSD typically used in studies relying on simulated data.  Notice that the design sensitivity curve is missing many important features found in real data, particularly the high-power, narrow-band features, or ``lines'', located throughout the sensitivity band of the detector.  Finally, the red (solid) curve shows a typical PSD that would be used during a parameter estimation follow-up analysis of a candidate chirp signal. To demonstrate the challenges of estimating the PSD using the methods employed in deriving the red curve, we will dissect the LIGO power spectrum into two components:  The broad-band noise as approximated by the design sensitivity curves, and the spectral lines found in the real data.

\subsection{The broadband LIGO/Virgo noise}

Ground based gravitational wave interferometer noise is dominated by three main components (details of the bandwidth for each effect depend on the instrument):  Sesimic noise steeply limits sensitivity below a few tens of Hz, thermal noise from the mirror suspensions and coatings dominate between the seismic wall and a few hundred Hz, and quantum (photon shot) noise is the limiting factor at high frequencies.  Each component is a broad-band effect with relatively simple behavior as a function of frequency, the sum of which produce the familiar sensitivity curves (e.g. the blue dotted line in Figure~\ref{fig:budget}) where the three different slopes are where the three different components -- seismic ($f\lesssim30$ Hz), thermal ($30\lesssim f\lesssim 100$ Hz), and shot noise ($f\gtrsim100$ Hz)-- dominate the noise budget.  The example shown in Figure~\ref{fig:budget} comes from 1024 s of data taken towards the end of LIGO's fifth science run -- when LIGO achieved initial design sensitivity -- which took place from November 2005 to October 2007, henceforth ``S5.''

Upgrades to the LIGO/Virgo facilities are, in a broad sense, intended to further suppress each of these three components.  A detailed discussions of the LIGO noise budget during S5 is found in Refs~\cite{LIGO-T050252, Kawabe:2008}.  For a preview of the expected noise budget for Advanced LIGO see Figure 2 of Ref.~\cite{Waldman:2011}.  Between the completion of S5, and before the full-scale commissioning work for Advanced LIGO began, the S6 science run took place between July 2009 to October 2010. S6 achieved better sensitivity than S5 but the power spectrum was qualitatively similar to Fig.~\ref{fig:budget}.  For the remainder of this article, all examples will use data taken during S6.

\begin{figure}[htbp]
   \centering
   \includegraphics[width=\linewidth]{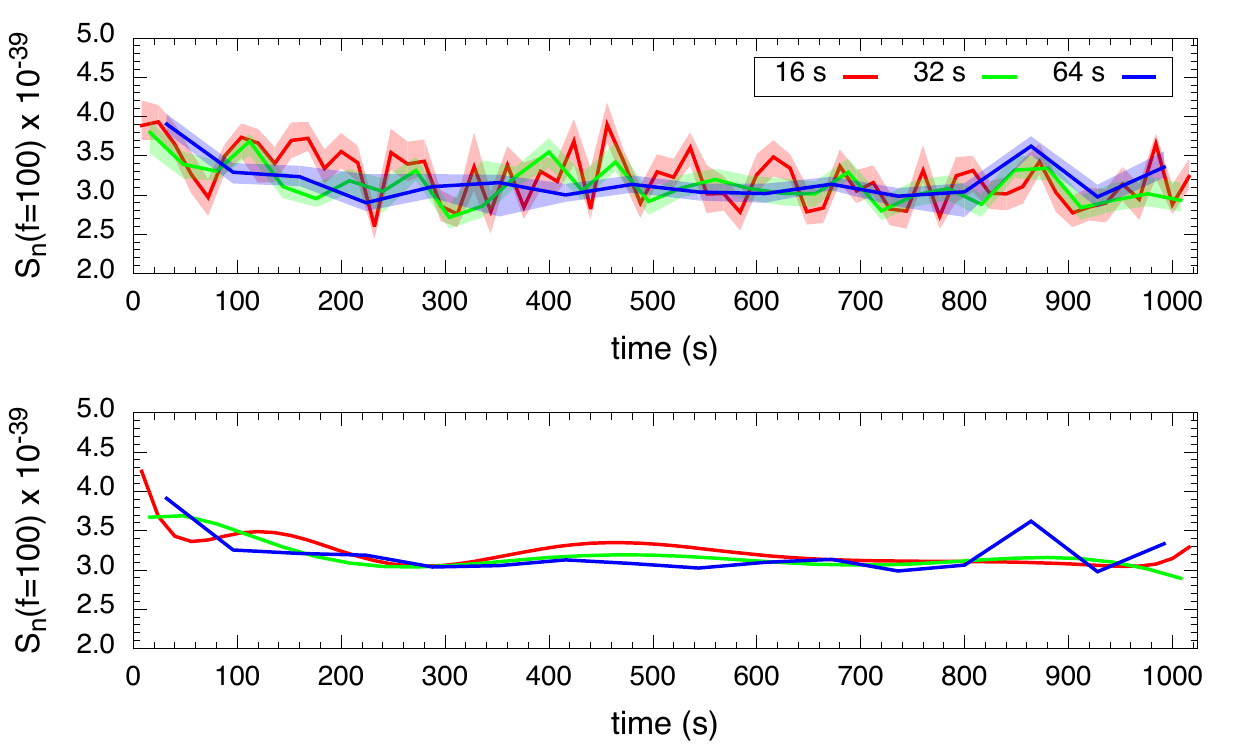} 
   \caption{\small {Demonstration of the time variability for the broad-band Gaussian noise.  Top panel:  PSD over 1024 s of S6 data at 100 Hz estimated using \BayesLine\  over intervals of 16 (red), 32 (green), and 64 (blue) seconds of data.  The thick line is the median PSD  from each segment's chain and the shaded region spans the $90\%$ credible interval.  Bottom panel:  Legendre polynomial fit to the data, where a RJMCMC was used to determine the best-fit degree ($n$) for the polynomial ($n=10$ for 16 s, $n=7$ for 32 s, and $n=11$ for 64 s ).}}
   \label{fig:legendre}
\end{figure}

To quantify how non-stationary the noise can be we begin with 1024 seconds of data which is the nominal duration used for PSD estimation in the follow-up of S6 triggers~\cite{S6PE,LALInference}.  We then divide that data into equal duration segments of width 32 s -- the amount of data which would be needed to perform parameter estimation of a binary neutron star signal in S6.  In each 32 second segment of data we use the \BayesLine\ algorithm (a detailed description of which can be found in Section~\ref{sec:BayesLine}) to compute $S_n(f)$ at 100 Hz, with 1$\sigma$ error bars as characterized by a Markov chain.  The PSD at 100 Hz is then plotted for each segment, as a function of segment start time relative to the full 1024 s of data, as the gray (dashed) points in  Figure~\ref{fig:legendre}.  We then repeat the analysis using 16 s  (red, solid) and 64 s  (blue dotted) seconds of data.  Notice that over the span of 1024 seconds -- the desired amount of data used to compute the average PSD during S6 -- we find statistically significant variation in the PSD.  To quantify the time dependence, we fit the time-dependent PSD to an $n^{\rm th}$-order polynomial.  The curves in Fig.~\ref{fig:bayesline} show the best-fit polynomials whose order $n$ ($10^{\rm th}$ for the 16 s segments, $7^{\rm th}$ for the 32 s segments, and $11^{\rm th}$ for the 64 s segments) had the highest Bayesian evidence as determined by a Reverse Jump Markov chain Monte Carlo (RJMCMC) code.  During the advanced detector era, a binary neutron star signal will be in band for this entire 1024 s segment of data and, using the same procedure as in previous analyses, nine hours of data would be required to estimate the PSD.  As can be seen from this demonstration, approximating the noise as being stationary over such long-duration intervals is not reliable.  By contrast, \BayesLine\ only needs the data being analyzed for a GW signal to determine the PSD, as opposed to requiring durations of data many times longer for spectral estimation.

\subsection{The spectral lines}
Above the broad-band noise floor for ground-based interferometric detectors are comparatively high power, narrow band, spectral features as seen in Figure~\ref{fig:budget}. The most significant spectral lines in Initial LIGO/Virgo data originate from one of three main processes:  Resonances of the wires which suspend the mirrors in the interferometer, henceforth referred to as ``violin'' modes, predominantly located around 320 Hz (plus higher harmonics) in initial LIGO; the AC electrical supply ``power line'' and subsequent higher frequency harmonics (at 60 Hz in the LIGO observatories); and calibration lines which are ``injected'' into the data by driving the mirrors at specific frequency and amplitude.  Advanced LIGO/Virgo will have similar features, though the central frequencies may differ. The precise frequency and amplitude of the environmental (power and violin) lines tend to drift on timescales shorter than the amount of data needed to estimate PSDs for in-spiral signals through averaging off-source data.  

\begin{figure}[htbp]
   \centering
   \includegraphics[width=\linewidth]{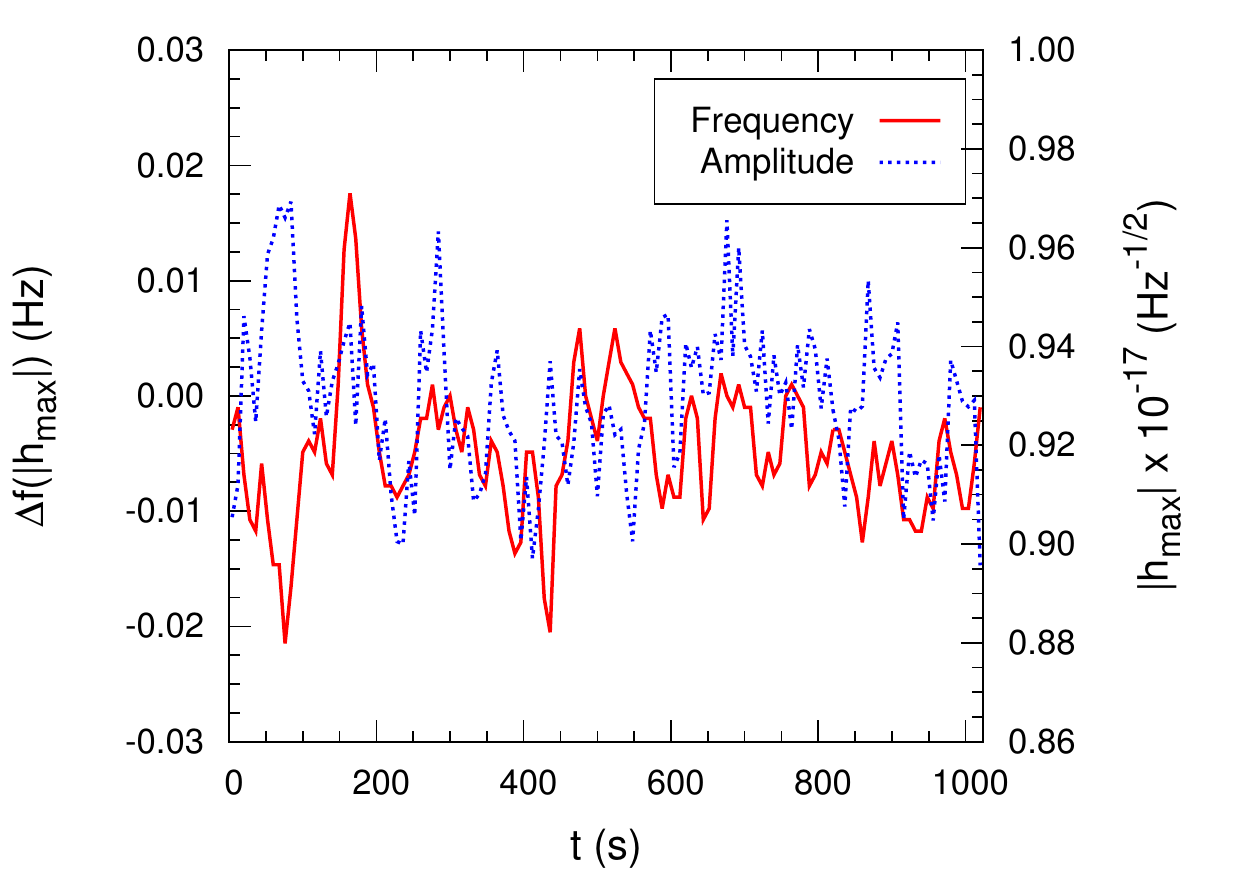} 
   \caption{\small {The 60 Hz power line shows non stationary frequency (left axis, red solid line) and amplitude (right axis, blue dashed line) over timescales shorter than the amount of data needed to characterize binary in spiral signals.  Higher frequency harmonics are practically 100\% correlated with the main power line.}}
   \label{fig:60Hz}
\end{figure}

Figure~\ref{fig:60Hz} depicts the time-evolution of the frequency (left axis, red solid line) and amplitude (right axis, blue dashed line) of the 60 Hz line computed by taking 1024 s of data and dividing it into 128 equal-length (8 second) segments.  Each segment of data was then Fourier transformed, and the frequency and amplitude of the peak in the power spectrum found in a window around 60 Hz was recorded.  Higher harmonics (e.g. at 120 Hz) show near-perfect correlation with the 60 Hz power line.

As is evident in Figure~\ref{fig:60Hz}, the amount of fluctuation in the location of spectral lines can be dramatic -- migrating by tens of Fourier bins for inspiral analyses -- which results in differences between the power spectrum used for the noise model and the actual power spectrum of the data.  

In principal, one could argue that notch-filtering the lines (i.e., excluding frequencies associated with line features from the likelihood integration in Eq.~\ref{eq:likelihood}) would be a simpler strategy.  However, what is not captured in Figure~\ref{fig:60Hz} are similar fluctuations in the line width.  Any notch filter would need to be conservative in order to ensure that no line power would leak into the integration domain and doing so would be at the expense of potentially throwing away good data.  A ``real time'' strategy for mitigating spectral lines is therefore preferred.

Several methods for dynamically removing lines have been put forward.  Analyses by Finn and collaborators using the 40 m prototype LIGO interferometer explored the use of Kalman filters~\cite{Finn:2000tt,LIGO-T060011} to subtract the violin modes from the data, while relying on auxiliary magnetometer data to approximate the contribution from the power lines in the gravitational wave channel and subsequently subtract the lines from the data.  Sintes and Schutz~\cite{Sintes:1998bu} used data from the Glasgow laser interferometer to develop the \emph{coherent line removal} algorithm which detects (and regresses) lines by searching for coherent harmonics of narrow-band features.

We will now introduce an alternative method for inferring $S_n(f)$ which is generically applicable to different spectral features in the data and does not depend on off-source data or long-term averages.  The algorithm is constructed with the end goal of simultaneously characterizing gravitational wave signals and detector noise in mind.  To do so we have created a framework by which parameters  for the Gaussian component of LIGO/Virgo noise can be seamlessly integrated into the existing parameter estimation software.  The algorithm, named \BayesLine\, uses RJMCMC methods to determine the most probable model for the PSD. What follows is a detailed description of the algorithm and examples designed to demonstrate its benefits for LIGO/Virgo data analysis

\section{The \BayesLine\ algorithm}\label{sec:BayesLine}
To infer the noise power spectral density from the data we have constructed a parameterized model for $S_n(f)$ which uses a MCMC to determine the posterior distribution function for the PSD.  MCMC methods have become increasingly common in astrophysics and are ubiquitous within the gravitational wave community, so in lieu of re-hashing the usual description of the algorithm we refer the reader to an accessible subset of the available literature~\cite{Metropolis:1953am, Hastings:1970, Cornish:2005qw, vanderSluys:2007st}. 

Just as our demonstrations of non-stationary LIGO noise separated the broad-band noise from the lines, we will model the two components separately, so that the total PSD $S_n(f)$ is the sum of the broad band ``smooth'' part of the power spectrum and the line model.  The dimension of the broad-band noise $S_{S}(f;\vec{\xi},N_S)$ and the spectral line models $S_L(f;\vec{\lambda},N_L)$ -- with  parameters $\vec{\xi}$ , $N_S$,   $\vec{\lambda}$, and $N_L$ to be defined in the next subsections -- is a free parameter, with the quantities $N_S$ (the number of control points in the cubic spline) and $N_L$ (the number of Lorentzians) tracking each component's dimension.  

For this application, $N_S$ and $N_L$ are nuisance parameters so we are not interested in selecting the PSD model with the largest evidence.  Instead, because we use an RJMCMC to sample the entire model space, we produce the model averaged distribution for $S_n(f)$.   As a result, inferences drawn from analyses which use \BayesLine\ will be marginalized over the noise model.

\subsection{The cubic spline model for broadband noise}
The smooth part of the PSD, $S_{S}(f;\vec{\xi},N_S)$, is fit using cubic spline interpolation. The 
parameters of the model are the number $N_S$, and location in frequency-PSD space $\vec{\xi}_i\rightarrow\left\{f_i, S_i\right\}$ of each spline point $i\in[0,N_S]$ in the interpolation.  
Between each pair of control points $[i,i+1]$  the noise spectrum is computed as a piece-wise polynomial by
\begin{equation}
S_{S}(f) = \sum_n^3 c_n^i\left(f-f_i\right)^n \ \text{for}\ f \in [f_i,f_{i+1}]
\end{equation}
where the coefficients ${\bf c}_n$ are found by using $\left\{\vec{\xi}_i\right\}$ as the table of points fed into the cubic spline routines available in the \emph{GNU Science Library}.

The spline control points start on a regular grid in frequency with an interval of $\sim10$ Hz and $S_{i}$ is initialized with the median Fourier power in a small window around frequency $f_i$.  During the Markov chain analysis the location of each spline point is adjusted both in frequency and power spectral density, existing points can be removed from the fit, and new points can be added.  

Priors for the spline model are uniform in frequency and in power spectral density.  The maximum number of control points is set by the total bandwidth of the data and the initial $\sim10$ Hz grid.  The range of allowed PSD values for each spline point depends on the type of data being analyzed, as we envision \BayesLine\ being a useful tool not only for gravitational wave data but also for auxiliary channels used to monitor the external environment at each observatory.

\subsection{The Lorentzian line model}
The most noticeable lines in the power spectrum are the ``violin modes'' caused by resonances in the mirror suspension system.  A good model for the source of the suspension lines is a noise-driven
and damped harmonic oscillators. The power spectrum for such systems is known to be a Lorentzian~\cite{Finn:2000tt} and so we will choose that as the basis for 
our model for the lines.  While the physical mechanism responsible for other spectral features found in the data -- including the 60 Hz  power line (plus harmonics) and calibration lines -- are not best represented by Lorentzians, we found that our choice of basis was adequate for all spectral features in the data and so opted for the simplest model to implement.  A more sophisticated model which is tailored for the different line-inducing processes in the data is discussed in the Appendix and left as a future direction to improve the \BayesLine\ algorithm.

The details of our Lorentzian line model are as follows:  We begin with a parameter vector 
$\vec{\lambda}_i\rightarrow\{A_i,f_i,Q_i\}$ for the $i^{\rm th}$ Lorentzian in the fit where 
$A_i$ is the amplitude, $f_i$ is the central frequency, and $Q_i$ is the quality factor which 
governs the width of the line.  

The full line model $S_{L}(f) $ is then the sum of each individual Lorentzian $\Lambda(f;\vec{\lambda}_i)$ in the fit
\begin{eqnarray}
S_{L}(f) &=& \sum_i^{N_L} \Lambda(f;\vec{\lambda}_i) \nonumber \\
\Lambda(f;\vec{\lambda}_i) &=& \frac{z(f)A_i f_i^4}{\left(f_i f\right)^2+Q^2(f_i^2-f^2)^2}
\end{eqnarray}
where $z(f)$ is used to truncate the tails of the Lorentzian distribution via
\begin{eqnarray}
z(f)= 
\begin{cases}
1,& \text{if } |f_i-f|\leq \delta f\\
e^{-\frac{f-\delta f}{\delta f}},              & \text{otherwise}
\end{cases}
\end{eqnarray}
with $\delta f\sim f_i/50$

Our priors for the central frequency of the spectral lines are uniform over the full bandwidth of the data and deserve further improvements:  Archived and auxiliary data provide information about the location of spectral lines; the power line (and its harmonics) are within a fraction of a percent of 60 Hz (and higher multiples);  calibration lines are purposefully added to the data at known frequency and amplitude;  and the suspension lines are mechanical resonances with narrowly constrained frequencies.   The ``signal to noise'' of the spectral lines is very high and will overwhelm our choice of prior, but making more informed selections about the spectral line parameters will improve the convergence of the \BayesLine\ model when integrated with algorithms which rely on parallel tempering for evidence calculations~\cite{BayesWave,LALInference}.

\subsection{The likelihood}
We arrive at the full noise power spectral density by adding the two components together
\begin{equation}
S_n(f) = S_L(f;\vec{\lambda},N_L) + S_{S}(f;\vec{\xi},N_S).
\end{equation}
To construct the likelihood we consider the joint probability that the $N$ complex Fourier coefficients of the data $\tilde{\vec d}$ would be realized from $S_n(f)$ assuming that the whitened data $\tilde{d}(f)/(\frac{T}{2}{S_n(f)})^{1/2}$ should be consistent with a unit normal distribution
\begin{equation}
\log p(\data|\vec{\lambda},N_L,\vec{\xi},N_S) = -\frac{2}{T}\sum_f^N\left[ \frac{|\tilde{d}(f)|^2}{S_n(f)}\right]
\end{equation}
where we have left off the constant normalization $-\frac{N}{2}\log(2\pi)$ as the MCMC is only concerned with relative likelihoods between points in parameter space.

\subsection{Reverse Jump Markov Chain Monte Carlo}

The central engine of \BayesLine\ is a trans-dimensional variant of a Markov chain -- RJMCMC -- where the chain transitions between different models for the data, potentially of different dimension, thereby simultaneously characterizing the posterior distribution function for each model and producing the evidence ratio, or Bayes factor, for each model~\cite{Green:2003}.  RJMCMC methods have been used in gravitational wave astronomy to assess wether simulated data contain a detectable astrophysical signal~\cite{Umstatter:2005jd, Cornish:2007if,Littenberg:2009bm, Stroeer:2009hj,Littenberg:2010gf,Sampson:2013lpa}, to estimate the mass distribution of stellar-mass black holes~\cite{Farr:2010tu}, and as a proof of principal for modeling non-Gaussian noise~\cite{Littenberg:2010gf}.  

RJMCMC has the unique ability to transition between competing models, effectively making the model one of the search parameters.  
Like its fixed dimension counterpart, the RJMCMC is guaranteed to (eventually) converge to the true target distribution -- the likelihood distribution across model space.  
The marginalized likelihood, or relative evidence, for each model is the number of iterations the chain spends in each model divided by the total number of iterations in the chain.  
A derivation of the error in RJMCMC evidence calculations is in our companion \BayesWave\ paper, Ref~\cite{BayesWave}.  

Allowing for the exploration of different models (which may differ in dimension) requires a separate Metropolis-Hastings step which proposes to move the chain from one model to another. 
Parameters $\pvec_i$ for trial model $\model_i$ are drawn from $q(\pvec_i|\model_i)$. 
If the models are nested, such as proposing additional parameters to include in the existing set, all of the like parameters are held fixed while the new parameters are drawn from $q$. 
Once the new models parameters are in hand the trans-model Hastings ratio is calculated by
\begin{equation}
\mathcal{H}_{\model_i\rightarrow\model_j} = \frac{   p(\data |\pvec_j,\model_j)  p(\pvec_j|\model_j) q(\pvec_i|\model_i)  }{   p(\data |\pvec_i,\model_i)  p(\pvec_i|\model_i) q(\pvec_j|\model_j)    } \vert J_{ij} \vert
\end{equation}
where the Jacobian $\vert J_{ij} \vert$ accounts for any change in dimension between models $\model_i$ and $\model_j$.
Selecting an efficient proposal distribution for model transitions is typically the major obstacle in the implementation of an efficient RJ routine. 
If the proposal distributions yield the model parameters directly, instead of a set of random numbers which are then used to determine the new model parameters, the Jacobian is unity and can be neglected. 

For the spline model uniform draws from the prior for the frequency and power spectral density of control points provide adequate mixing of trans-dimensional proposals, though this would improve with stricter priors.  For the line model we again rely on uniform draws from the prior for the amplitude and quality factor for the Lorentzians.  For the central frequency the power spectrum is first divided into several narrow-band segments which are weighted by the power in that segment divided by the median power of the full data, thereby preferring to propose Lorentizians in segments that have a large excess of power.  Segements are proposed according to their relative weighting and the frequency is drawn uniformly from within the narrow-band segment.  We verify that detailed balance is satisfied by running \BayesLine\ with the likelihood set to a constant and confirming that the recovered posterior agrees with our priors.

\section{Studies using LIGO data from S6} \label{sec:S6}
With the likelihood and prior distributions defined, we can now completely characterize the posterior distribution function of the PSD using the RJMCMC algorithm.
We will now make use of this formalism to demonstrate how our parameterized PSD outperforms the off-source averaging method for spectral estimation using 
data acquired by the LIGO observatories during S6.

For the stopping condition of the Markov chain we determine when the chain has converged by using a posterior predictive check which demands that the whitened data (which should be consistent with a normal distribution) do not have large-sigma outliers. We do not currently monitor the autocorrelation length to adaptively determine when enough posterior samples have been collected after the convergence criterion is satisfied, instead using a fixed number of samples in the chain post-convergence that has proven to work well.  As \BayesLine\ is incorporated into existing LIGO/Virgo Bayesian inference pipelines the stopping criteria will need to be more carefully handled.

Run-times for the current \BayesLine\ implementation are a factor of a few times larger than the duration of data being analyzed, depending on the sampling rate.  For example, \BayesLine\ requires $\sim 1$ hour to estimate the PSD for 1024 seconds of data sampled at 4096 Hz.  The run time is insignificant when compared to the computational cost of compact binary parameter estimation analyses~\cite{LALInference}.

\begin{figure}[htbp]
   \centering
   \includegraphics[width=\linewidth]{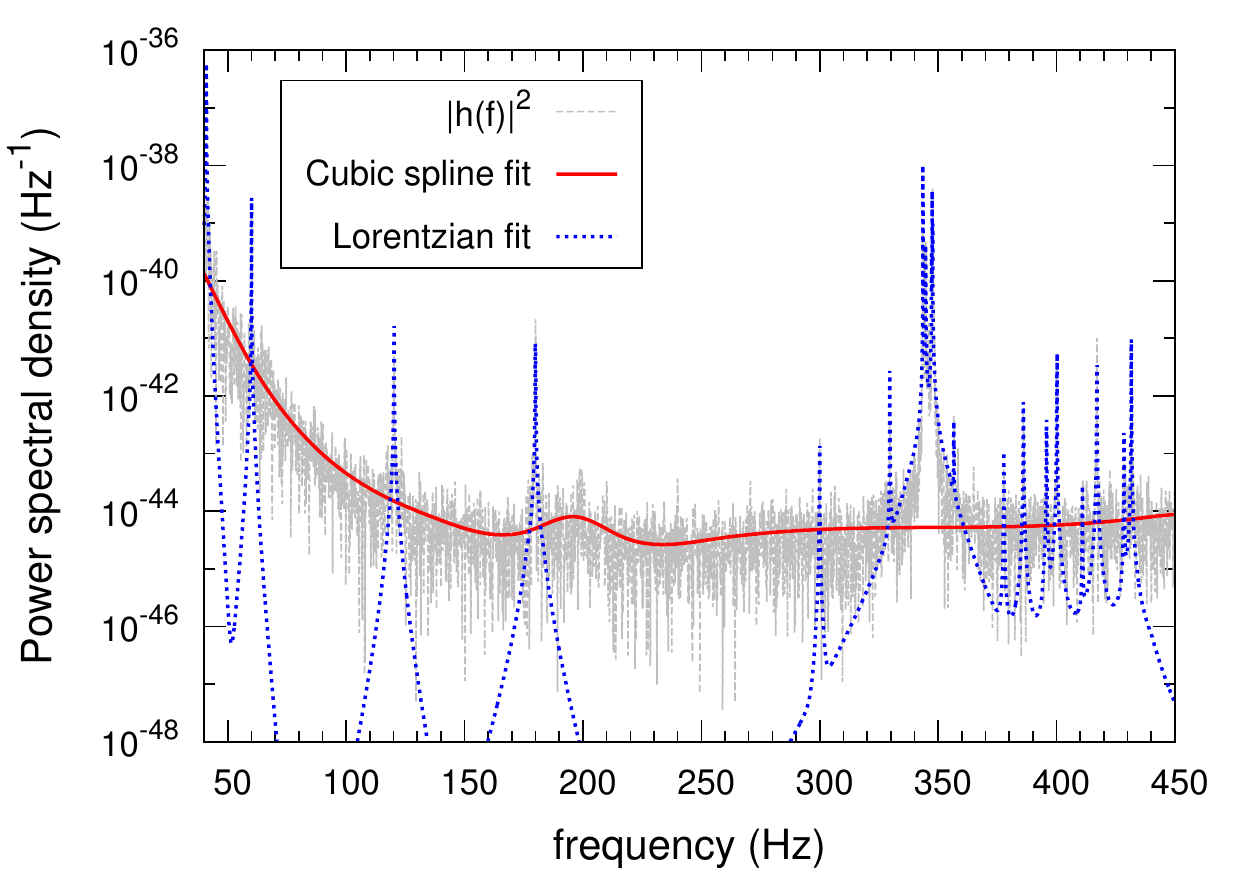} 
   \caption{\small {Example S6 data power spectrum (gray, dashed) with \BayesLine\ PSD shown separated into the cubic spline (red, solid) and Lorentzian (blue, dotted) components.}}
   \label{fig:bayesline}
\end{figure}

To begin we show in Figure~\ref{fig:bayesline} a typical S6 LIGO power spectrum (gray, dashed) and and the two component \BayesLine\ maximum a posteriori PSD (with the appropriate $T/2$ normalization).  The spline model is shown in the red (solid) curve while the Lorentzians used to fit the spectral lines are depicted by the blue (dotted) lines.  The full PSD would be sum of the spline and Lorentzian fit.  For this example we use $T=32$ s of data from the LIGO Livingston Observatory.  Our choice for $T$ is consistent with the amount of data required for parameter estimation follow up of a binary neutron star (BNS) signal.  BNS in-spirals pose the most stringent challenge for PSD estimation (among the transient sources) because the duration of the signal increases with decreasing mass. Binary neutron stars are anticipated to be the lowest mass signals and, as a result, require the longest data segments. 

Using averaged off-source data to estimate the PSD would demand $\gtrsim 1024$ s of data.  To show the difference between the \BayesLine\ noise spectrum and a PSD estimated by averaging we use the 1024 seconds immediately after the 32 s segment of data used in Figure~\ref{fig:bayesline} to compute $S_n(f)$ as was done during S6.  Had the data from Figure~\ref{fig:bayesline} contained a candidate detection, this would be indicative of the PSD used in the analysis.  To compare the two PSDs we use each of them to whiten the original 32 s of data, and then histogram the real and imaginary Fourier coefficients.  

\begin{figure}[htbp]
   \centering
   \includegraphics[width=\linewidth]{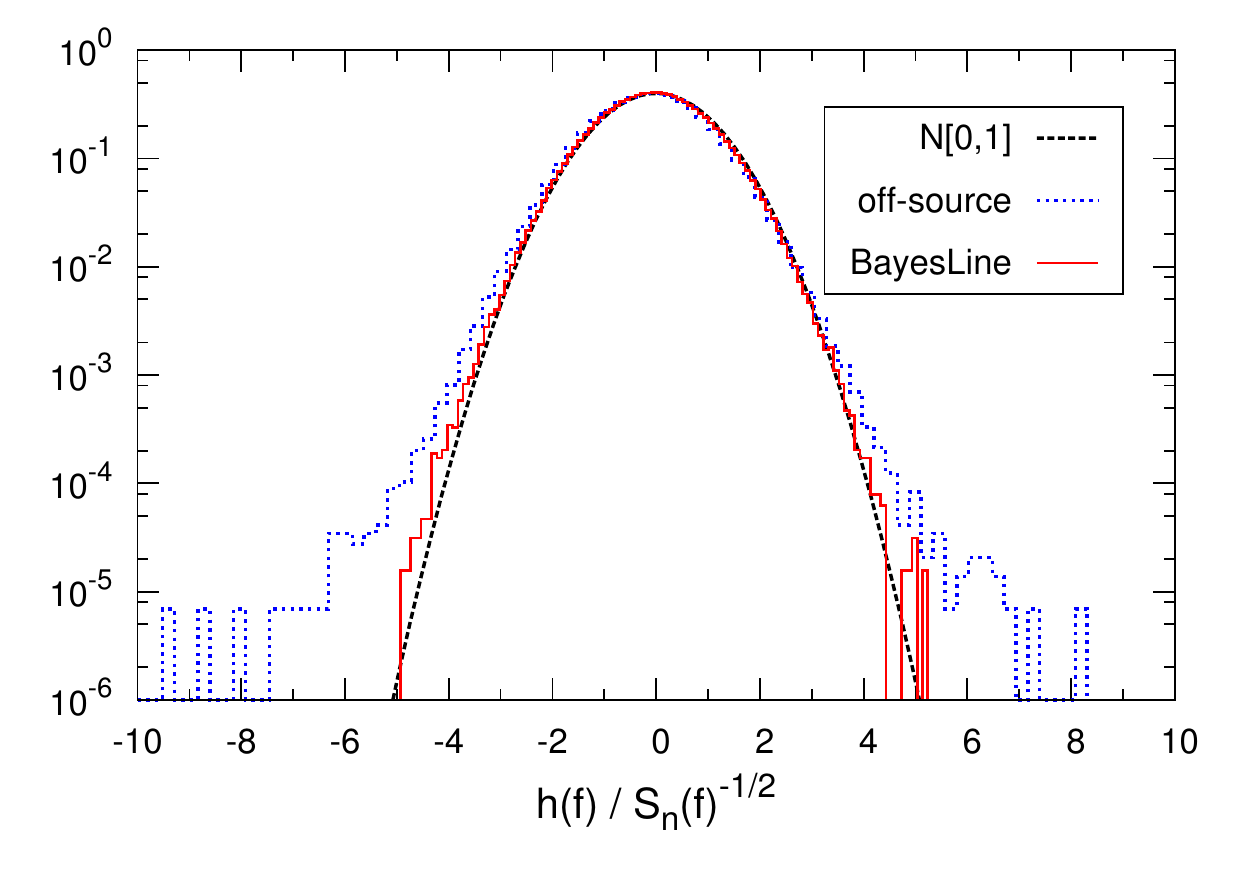} 
   \caption{\small {Initial LIGO example distribution of whitened Fourier domain data using the standard PSD estimation in \LALInference\ (blue, dotted) and the \BayesLine\ PSD (red, solid), compared to a normal distribution with zero mean and unit variance (gray, dashed) which is assumed by the likelihood function.}}
   \label{fig:eLIGOWhite}
\end{figure}

According to the likelihood function used for parameter estimation (Eq.~\ref{eq:likelihood}) the whitened data should be drawn from a normal distribution with zero mean and unit variance.  Figure~\ref{fig:eLIGOWhite} shows the distribution of the whitened data using the \BayesLine\ PSD (red, solid) and the off-source PSD (blue, dotted) as compared to a zero-mean unit-variance normal distribution $N[0,1]$ (gray dashed).  The difference is striking -- using the off-source PSD leaves behind significant large $\sigma$ tails.  Because the likelihood is maximized when the whitened residual looks Gaussian, the excess tails left behind by the off-source whitening invite bias by the gravitational wave model in an attempt to account for the non-Gaussian residual.  On the other hand, our \BayesLine\ PSD produces a Gaussian residual thereby suppressing the potential for large systematic errors in GW parameter estimation due to the PSD estimation.

\begin{figure}[htbp]
   \centering
   \includegraphics[width=\linewidth]{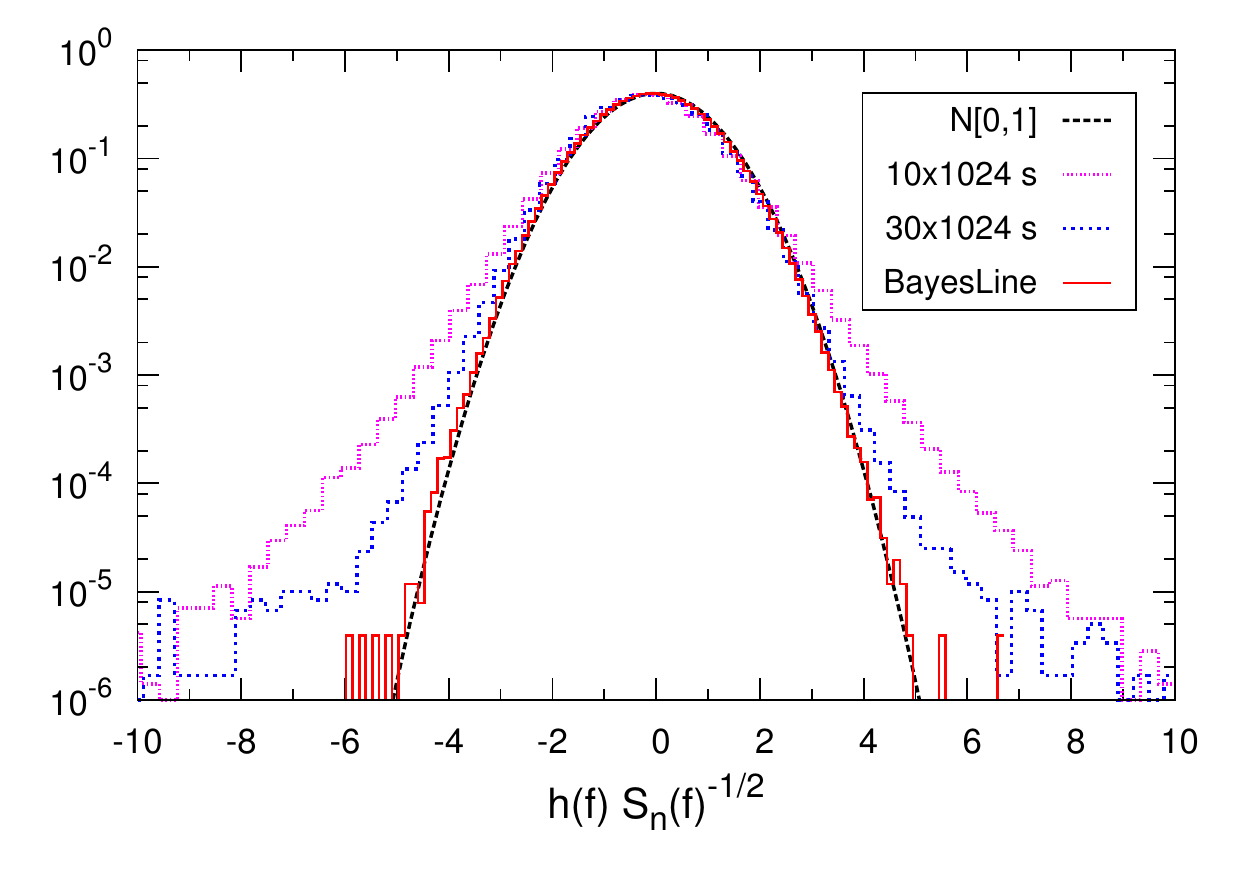} 
   \caption{\small {Example distribution of whitened Fourier domain data made from the amount of data consistent with what is needed for Advanced LIGO BNS analyses.  PSDs were determined via the standard PSD estimation in \LALInference\ using 10 (magenta, dotted) and 30 (blue, short dashed) 1024 s segments of data for averaging and the \BayesLine\ PSD (red, solid).  All three distributions are compared to $N[0,1]$ (gray, long dashed).}}
   \label{fig:aLIGOwhite}
\end{figure}

As motivated in Sec.~\ref{intro}, estimating $S_n(f)$ for long-duration signals in the advanced detector era using off source data is not feasible.  To illustrate this reality we repeat the analysis from Fig.~\ref{fig:eLIGOWhite} but using observation times $T=1024$ s consistent with a BNS signal in Advanced LIGO/Virgo.  We estimate the PSD and histogram the whitened Fourier coefficients, shown in Figure~\ref{fig:aLIGOwhite}, using \BayesLine\ (red, solid) $10\times1024$ (magenta, small dots) and $30\times1024$ (blue, large dots) seconds of off-source data.  The 10240 seconds we consider the bare minimum for PSD estimation, while the $30720$ s segment uses the same number of segments (30) in the Welch average as the S6 example for advance LIGO-like segment lengths.  Note that $\sim$30000 seconds ($\sim$8 hours) of stationary noise is unlikely to be available due to environmental disturbances affecting data quality.   


\begin{figure}[htbp]
   \centering
   \includegraphics[width=\linewidth]{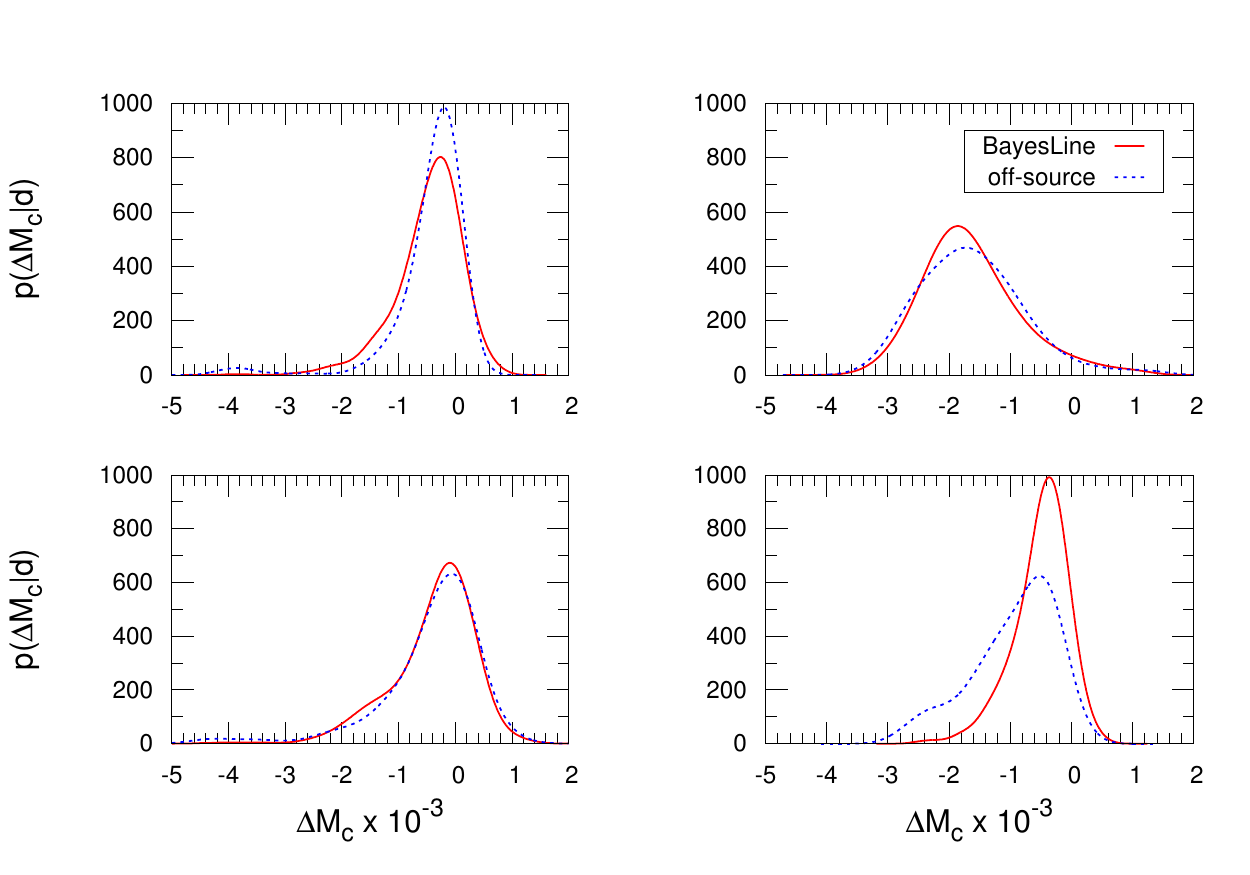} 
   \caption{\small {Posterior distribution functions from four different simulated signals in real detector noise for the chirp mass of a binary neutron star recovered using \LALInference\ with the \BayesLine\ PSD (red, solid) and the Welch-averaged off-source PSD (blue, dotted).  The $x$-axis is the difference between the recovered and the injected chirp mass, so $x=0$ corresponds to the true value.  Using the \BayesLine\ PSD does not degrade parameter recovery and statistically significant differences are found in the posteriors. }}
   \label{fig:posterior}
\end{figure}

To demonstrate how the PSD model can impact parameter estimation we add a simulated BNS signals to the data and then use the LIGO/Virgo parameter estimation software \LALInference\ to recover the signals and characterize their posterior distribution functions~\cite{LALInference}.  We add a signal to each of the 32 s sub-segments of the full 1024 s of data which has been under scrutiny throughout this work and compare the results when using the PSDs from Figure~\ref{fig:eLIGOWhite}.   Figure~\ref{fig:posterior} shows the posterior distributions for the chirp mass $M_c = (m_1m_2)^{3/5}/(m_1+m_2)^{1/5}$, where $m_1$ and $m_2$ are the masses for the two neutron stars in the binary, for four of the signals injected into the different 32 s data segments.  The $x$-axis shows $\Delta M_c = M_c - M_c^{\rm true}$ so that zero corresponds to the injected value.  The red (solid) distributions use the \BayesLine\ PSD while the blue (dotted) distributions Welch average off-source data.  We find that, for some cases, the two agree quite well (bottom left and top right) but for others the difference between the two distributions is significant.

\begin{figure*}[htbp]
   \centering
   \includegraphics[width=\linewidth]{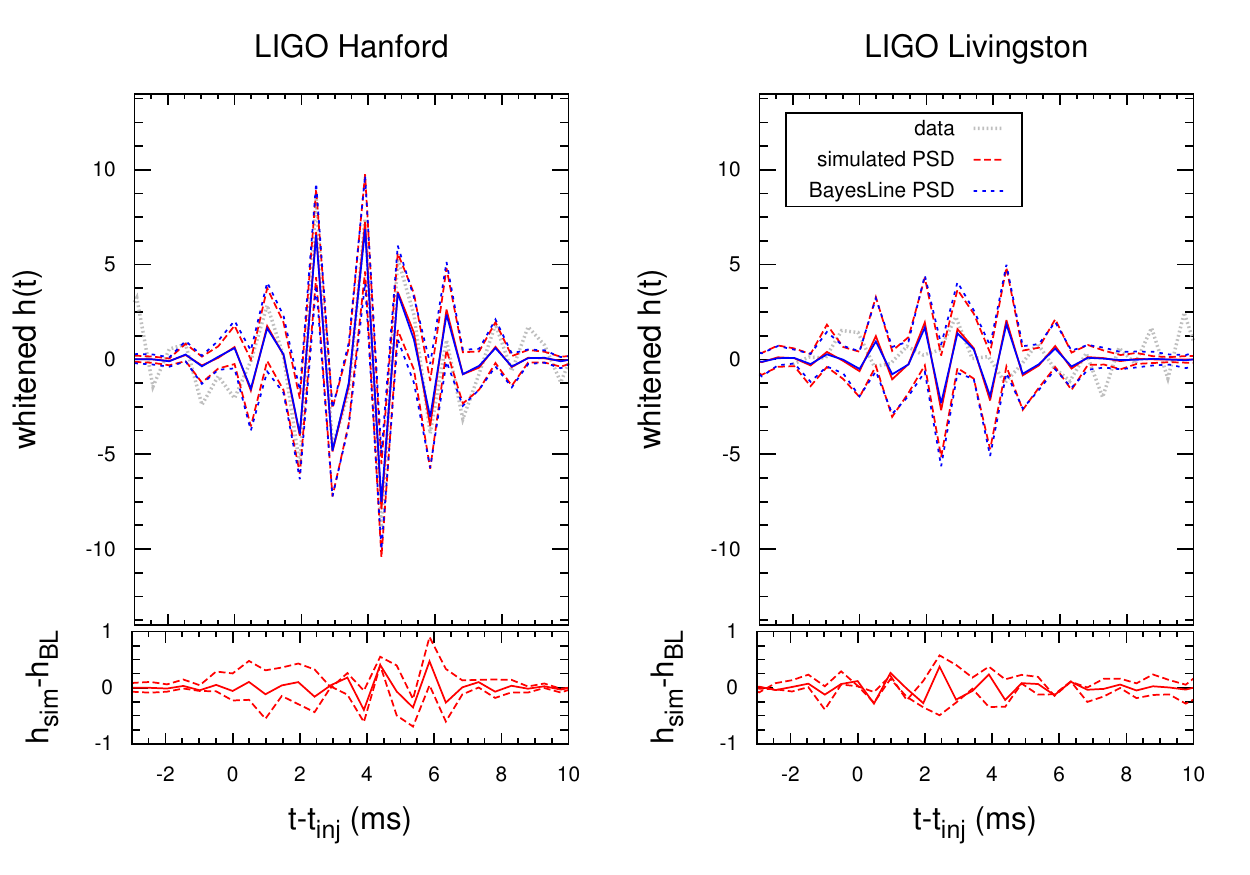} 
   \caption{\small {Comparison of reconstructed waveforms in simulated data using the true PSD (red) and the \BayesLine\ PSD (blue).  The top panels are the whitened data (grey) and waveforms.  Solid lines indicate the median reconstructed signal while dashed lines span the $1\sigma$ errors.  The bottom panel shows the whitened residual between the two recovered waveforms, which is consistent with zero.  The figure demonstrates how including the \BayesLine\ PSD parameters does not degrade signal recovery.}}
   \label{fig:wave}
\end{figure*}

In principle there is a risk that the PSD model is able to fit-out part of the gravitational wave signal and thus introduces, instead of resolves, biases.  In practice this is not a problem.  Our ultimate vision of having \BayesLine\ and gravitational wave parameters simultaneously in the model will not suffer from this hazard because the template provides a \emph{much} better model for the GW signal than \BayesLine\ and so the RJMCMC will, by construction, prefer models which do not interfere with the signal characterization.  We demonstrate this point using simulated data (so the true PSD is known) and a simple model for the gravitational wave signal and template (a sine Gaussian).  The data is analyzed once using the LIGO design sensitivity curve (which was used to simulate the noise), and again using \BayesLine\ and the GW model simultaneously.  We show the agreement between our fit and the true PSD, as well as between the recovered waveforms from the two runs in Figure~\ref{fig:wave}.  The top panels show the whitened simulated data in gray, and the recovered waveforms are plotted using the true PSD (red) and the  \BayesLine\ PSD (blue).  The solid lines are the median waveform and the dotted lines span the $1\sigma$ errors for the reconstructed waveform.  The agreement between the recovered waveforms using the different PSDs is excellent.  The bottom panel shows the median and 1$\sigma$ residuals between the two reconstructed waveforms, which are consistent with zero.

\section{Conclusions and future goals} \label{sec:Conclusions}
Analysis of ground-based gravitational wave interferometer data is challenging in part because the instrument noise, over intervals of time longer than duration of transient signals, is not stationary making spectral estimation a challenge.  The standard approach to noise characterization used by analyses for binary in-spiral signals involves averaging over long durations of data compared to the amount of time that the signal is in band and is thereby susceptible to systematic errors.  In this paper we provide an alternative approach to PSD estimation -- dubbed \BayesLine\ -- for inferring the noise spectrum directly from the data being analyzed.  Doing so within the MCMC framework means that our inferences about GW detections can be marginalized over all PSDs consistent with the data.  The \BayesLine\ algorithm has the added benefit of freeing the analyses from needing to use large amounts of data for spectral estimation.  

 \BayesLine\ has been developed outside of any existing LIGO/Virgo analysis pipelines and, for this work, is used as a stand-alone tool for PSD estimation.  We have described the algorithm and demonstrated its ability to produce estimates to LIGO noise spectra which offer advantages over methods historically used for binary in-spiral parameter estimation analyses.  \BayesLine\ only addresses the stationary, Gaussian component of the noise and is otherwise insensitive to glitches which have caused problems for the analysis both in regards to false alarm rates and parameter estimation biases.  The companion algorithm which does model transient non-Gaussian noise has been combined with \BayesLine\ and is described in Ref~\cite{BayesWave}.

The \BayesLine\ method will reach its full potential when it is integrated into other data analysis pipelines.  We will next look to incorporate our spectral estimation method into the \LALInference\ pipeline for parameter estimation follow up of binary in-spiral and merger signals.  

The priors used in \BayesLine\ are very wide and pose challenges for computing evidence integrals.  A high priority improvement is to develop more informative priors which will improve efficiency of the RJMCMC and make evidence integrals tractable.  One method currently under consideration is to construct kernel density estimates of the PSD using archived data.  In this paradigm, \BayesLine\ would run continuously on data as it is collected and continually update the kernel density estimate used for the prior in follow-up analysis of candidate signals.

\BayesLine\ still implicitly assumes that the noise in the segment of data being analyzed is stationary which we have demonstrated not to be true for the low-mass compact binary signals.  
We envision developing a version of \BayesLine\ that works
across multiple small segments, with the parameters in each segment linked together by using
the spectrum from neighboring segments as a prior. 

Our model for the PSD is effective but is completely phenomenological.  A physically motivated model for the broad-band noise and spectral lines will provide more information about the detector.  For the smooth broad-band contribution to the PSD a physical model for the quantum noise can be found in Ref~\cite{Buonanno:2001cj} and phenomenological models for seismic and thermal noise have been constructed -- all of which are parameterized by quantities relating to the detector.   A more physically motivated model for the lines is also an intriguing possibility.  For example, the 60 Hz line invites coherent modeling and subtraction from the data, as opposed to whitening.  Doing so will provide a cleaner residual for long-duration data segments and could also provide improved efficiency because harmonics may not need to be independently modeled.  We conducted a preliminary study when deciding between regressing and whitening the spectral lines the results of which can be found in the Appendix.  


\section{Acknowledgements}
We would like to acknowledge hospitality of the Center for Interdisciplinary Exploration and Research in
Astrophysics (CIERA) at Northwestern University where the \BayesLine\ algorithm was, in part, conceived.  Will Farr and Vicky Kalogera participated in the early discussions that resulted in the \BayesLine\ algorithm and have provided suggestions throughout its development.  We thank Michael Coughlin, Yiming Hu, and Matthew Pitkin for helpful comments on an earlier draft of this paper.  NJC appreciates the support of NSF Award PHY-1306702. TBL acknowledges the support of NSF Award PHY-1307020.

\bibliography{/Users/tyson/Research/Papers/papers}

\appendix
\section{Subtracting the power lines}\label{sec:60Hz}

Coherently modeling and regressing spectral lines from the data, as is done for signals/glitches~\cite{BayesWave}, is an alternative to the whitening approach taken here.  Instead of modeling the \emph{power} contained in the spectral lines -- appearing in the denominator of Eq.~\ref{eq:likelihood} -- we can include the time-dependent \emph{phase} and \emph{amplitude} evolution of the spectral lines in our calculation of the residual in the numerator.  For demonstrative purposes we will focus our attention on the 60 Hz power line found in the LIGO data for which a Lorentzian is less-well physically motivated as the line profile.

If the 60 Hz line had constant frequency it would put in a single spike in the Fourier series of the data, spread due to how the electronics are coupled to the strain measurement and finite sampling effects, which would be well represented by a sinusoid.  Temporal evolution will further broaden the spectral lines in \emph{a priori} unpredictable ways making sinusoids a convenient basis to work in, but necessitating another adaptive-dimension linear combination to provide a parsimonious fit to the data.

We experimented with modeling the power lines as sinusoids with amplitude $A(t)$ and frequency $f(t)$ which we compute in the time-domain, Fast Fourier Transform (FFT) the time-domain model, and then subtract from the frequency-domain data.   The functions $A(t)$ and $f(t)$ are modeled by a sum of sines and
cosines. The number of sines and cosines used in each model again determined by a RJMCMC. 

\begin{figure}[htbp]
   \centering
   \includegraphics[width=\linewidth]{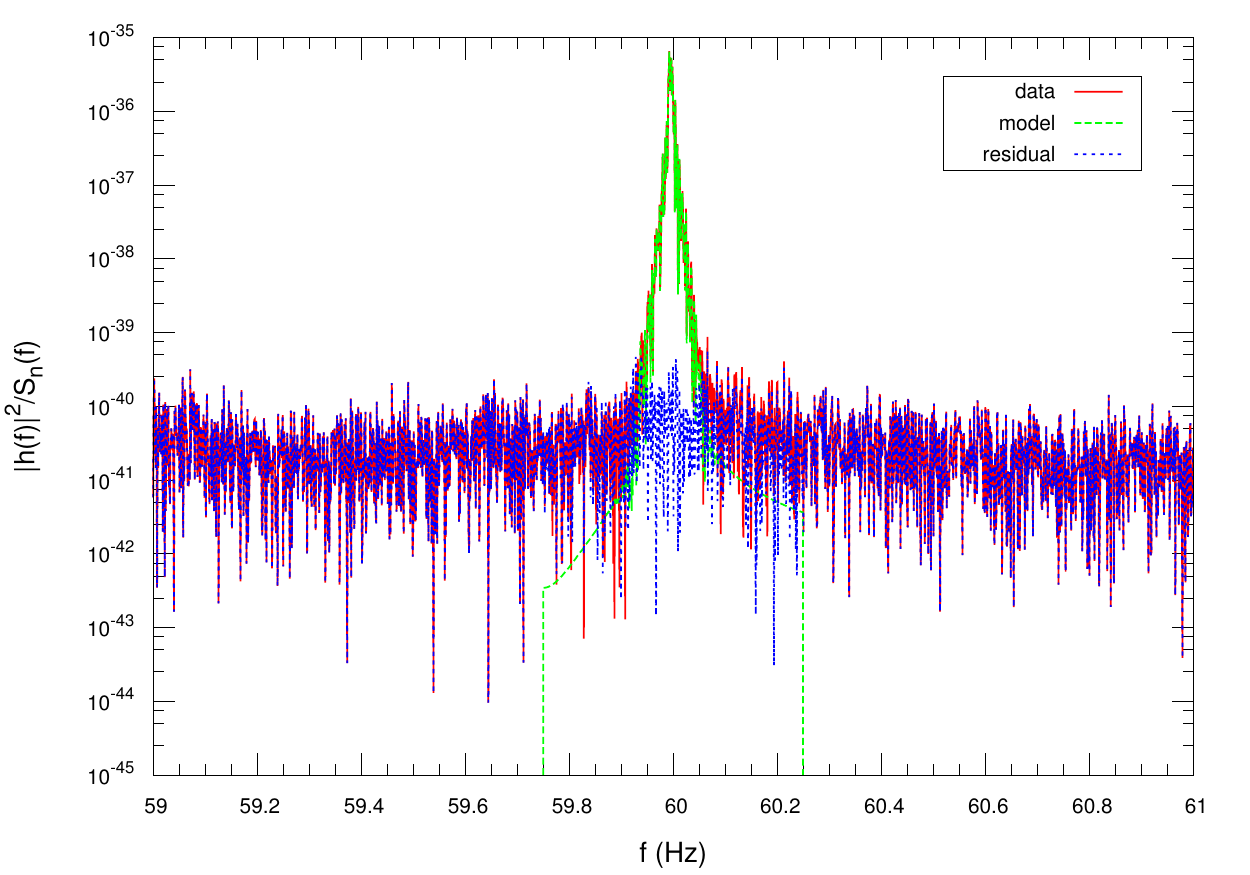} 
   \caption{\small {The power spectrum of the data [red, solid], the coherent 60 Hz line model [green, dashed] and the residual [blue, dotted] after the line is subtracted.}}
   \label{fig:60HzModel}
\end{figure}

\begin{figure}[htbp]
   \centering
   \includegraphics[width=\linewidth]{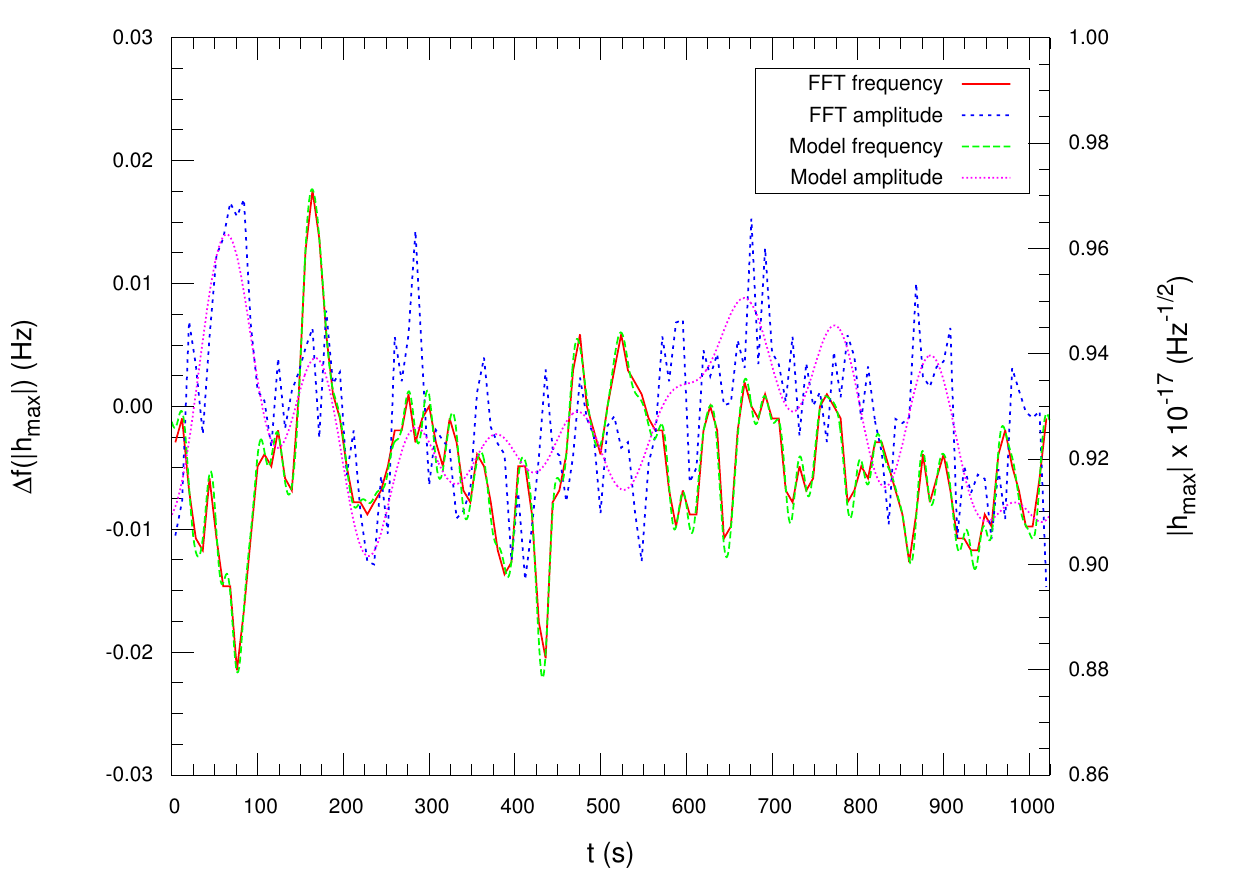} 
   \caption{\small { The frequency [left axis; red, solid] and amplitude [right axis; blue short-dashed] of the data as a function of time as estimated by taking FFTs of each 16 s interval.  The best fit model for $f(t)$ and $A(t)$ derived from the full 1024 seconds of data are shown in the green long-dashed curve and the magenta dotted curve, respectively.}}
   \label{fig:drift2}
\end{figure}

To test the coherent approach we analyzed 1024 s of LIGO Livingston S6 data.  We found the $f(t)$ required $\sim 60$ basis functions, while  $A(t)$ was easier to model, using $\sim 10$.  Figure~\ref{fig:60HzModel} shows the power spectrum of the data [red, solid], the coherent 60 Hz line model [green, dashed] and the residual [blue, dotted] after the line is subtracted.  Figure~\ref{fig:drift2} shows the frequency [left axis; red, solid] and amplitude [right axis; blue short-dashed] of the power spectrum's peak as a function of time as estimated by Fourier transforming each 16 second interval within the full 1024 s data segment.   Our best fit model for $f(t)$ and $A(t)$ analyzing the full 1024 seconds of data are shown in the green long-dashed curve and the magenta dotted curve, respectively. The remarkable thing about our result is the agreement between the short FFT estimates for $f(t)$ and the RJMCMC estimate for the best $f(t)$ to subtract the line from the data. These are two totally independent ways of estimating the frequency evolution of the 60 Hz line.


The sum-of-sinusoids approach worked well for the power lines but other narrow-band features in the data, especially lines from the mirror suspension system, required a challenging number of basis functions.
As discussed in the text, the suspension lines are noise driven and damped harmonic oscillators.  The power spectrum for such systems is known to be a Lorentzian.  We the sum-of-Lorentzians model better handled the suspension lines and did a fine job with the power and calibration lines as well. Rather than have a complicated multi-component model for the different types of lines we opted for a more simplified approach.


Note that unlike the Lorentzian model, which does power spectrum whitening, the power line model does coherent
subtraction.  The line-subtraction approach can be adapted to simultaneously remove harmonics of the 60 Hz line using a single model for $f(t)$ and $A(t)$ plus an overall scale factor for the amplitude of each harmonic.  Early indications of this approach were promising but never reached maturity because the preliminary study was done before the cubic-spline model for the broad-band noise was developed.  The multi-harmonic line model could potentially reduce the overall dimension of the line model which would in turn simplify the MCMC implementation.

We will continue to develop a coherent model for the power lines, incorporating environmental monitoring channels at the observatories to more directly measure the behavior of the electrical supply. 

\end{document}